\documentstyle[12pt]{article}

\begin{document}

\title{A simple explanation of the non-appearance \\ of physical gluons and quarks}
\author{Johan Hansson \\ Department of Physics \\ Lule{\aa} University of
Technology \\ SE-971 87 Lule\aa, Sweden \\ \texttt{E-mail:
hansson@mt.luth.se}
\vspace{0.5cm} }

\date{}

\maketitle

\begin{abstract}
We show that the non-appearance of gluons and quarks as physical
particles is a rigorous and automatic result of the full, {\it
i.e.} nonperturbative, nonabelian nature of the color interaction
in quantum chromodynamics. This makes it in general impossible to
describe the color field as a collection of elementary quanta
(gluons). Neither can a quark be an elementary quantum of the
quark field, as the color field of which it is the source is
itself a source, making isolated noninteracting quarks, crucial
for a physical particle interpretation, impossible. In geometrical
language, the impossibility of quarks and gluons as physical
elementary particles arises due to the fact that the color
Yang-Mills space does not have a constant trivial curvature.

In QCD, the particles ``gluons'' and ``quarks'' are merely
artifacts of an approximation method (the perturbative expansion)
and are simply absent in the exact theory. This also coincides
with the empirical, experimental evidence.
\end{abstract}

\newpage
One major problem in contemporary particle physics is to explain
why quarks and gluons are never seen as isolated particles. A lot
of effort has gone into trying to resolve this puzzle over a
period of years. Different approaches include lattice QCD, dual
Meissner effect (in the QCD-vacuum), instantons, etc, but the
problem is not yet fully solved. For a review, see \cite{Bander}.

We will take a different route than normally used, to eliminate
the problem before it arises.

Usually, most particle physicists use ``fields'' and ``particles''
interchangeably, {\it i.e.} as denoting the same things. That is
because the almost universal usage of Feynman diagrams gives the
false impression that particles (quanta)  are always exchanged,
even when they cannot exist. This is an example of mistaking the
approximation (perturbation theory) for the exact theory. However,
this misconception seems to be so common that many physicists do
not even note, or care about, the distinction. The usage of the
Feynman diagram technique (propagators, Green's functions, etc)
can be justified as an approximation for mildly nonlinear theories
(weak coupling limit) but breaks down for strongly coupled
nonabelian theories. (And also for strongly coupled abelian
theories with {\it sources}.)

In quantum chromodynamics (QCD) it is at first
sight a puzzle why the color force should be short-range, and
especially why gluons are not seen as free particles, as the
nonbroken SU(3) color symmetry {\it seems} to demand massless
quanta, which naively would have infinite reach. However, as we
shall see, there generally {\it are} no quanta.

In quantum field theory an {\it elementary particle}
\cite{BjorkenDrell}, {\it i.e.} a {\it quantum} $=$ a harmonic
excitation of a fundamental field \cite{Feynman}, is defined
through the creation and annihilation operators, $a^{\dagger}$ and
$a$, of the ``second-quantized" theory. For instance, in quantum
electrodynamics (QED), the entire electromagnetic field can be
seen as a collection of superposed quanta, each with an energy
$\omega_k$. The hamiltonian of the electromagnetic field (omitting
the zero-point energy) can be written

\begin{equation}
H  =  \sum_k N_k \omega_k,
\end{equation}
where

\begin{equation}
 N_k = a_k^{\dagger} a_k ,
\end{equation}
is the ``number operator'', {\it i.e.} giving the number of quanta
with a specific four-momentum $k$  when operating on a free state,

\begin{equation}
 N_k |... n_k ... \rangle = n_k |... n_k... \rangle.
\end{equation}
As all oscillators are independent,

\begin{equation}
 |... n_k ... \rangle = \prod_k  | n_k \rangle,
\end{equation}
where $n_k$ is a positive integer, the number of quanta with that
particular momentum. The energy in the electromagnetic field is
thus the eigenvalue of the hamiltonian (1). The reasoning for
fermion fields is the same, but then the number of quanta in any
given state can be only 0 or 1 (``Fermi statistics").

Now, assuming that QCD is the correct theory of quark
interactions, a problem arises, as it is generally impossible to
write the color fields in terms of superposed harmonic
oscillators. It is not possible to represent the solution as a
Fourier expansion and then interpret the Fourier coefficients as
creation/annihilation operators through ``second quantization", as
the color vector potentials $A_{\mu} ^b$ ($b \in 1,...,8$) are
governed by \textit{nonlinear} evolution equations,
\begin{equation}
D^{\mu} F_{\mu \nu} = j_{\nu},
\end{equation}
and Fourier methods are inapplicable to nonlinear equations (see,
\textit{e.g.}, \cite{Zwillinger}).

Without a quark current, $j_{\nu} \equiv g_s \bar{\psi}
\gamma_{\nu} \psi = 0$, we get, in component form

\begin{eqnarray}
(\delta_{ab} \partial^{\mu} + g_s f_{abc} A_c^{\mu})
(\partial_{\mu} A_{\nu}^b - \partial_{\nu} A_{\mu}^b + g_s f_{bde}
A_{\mu}^d A_{\nu}^e) = 0,
\end{eqnarray}
where $g_s$ is the color coupling constant (summation over repeated
indices implied).

When we have an abelian dynamical group, as in QED, all the
structure constants $f_{abc}$  are zero. Eq.(6) then reduces to a
linear differential equation, and a general solution can be
obtained by making the Fourier expansion:
\begin{eqnarray}
A_{\mu}^{QED} = \int d^3 k \sum_{\lambda = 0}^3 [a_k (\lambda)
\epsilon_{\mu}(k, \lambda) e^{-i k \cdot x} + a_k^{\dagger}
(\lambda) \epsilon_{\mu}^{\ast} (k, \lambda) e^{i k \cdot x}],
\end{eqnarray}
where $\epsilon_{\mu}$ is the polarization vector. We have also
omitted an, for our purposes, inessential normalization factor.

However, for a theory based on a nonabelian group, like QCD, this
can no longer be done \cite{Zwillinger}, due to the nonlinear
nature of Eq.(6) when $f_{abc}, f_{bde} \neq 0$,
\begin{eqnarray}
A_{\mu}^{b \, QCD} \neq   \int d^3 k \sum_{\lambda = 0}^3 [a_k^b
(\lambda) \epsilon_{\mu}(k, \lambda) e^{-i k \cdot x}  + a_k^{b \,
\dagger} (\lambda) \epsilon_{\mu}^{\ast} (k, \lambda) e^{i k \cdot
x}].
\end{eqnarray}

Thus, the color fields can be represented by harmonic oscillators
(gluons) only in the trivial, and physically empty, limit when the
strong interaction coupling constant tends to zero, $g_s
\rightarrow 0$ (or, within perturbation theory, equivalently when
$Q^2 \rightarrow \infty$ because of asymptotic freedom). Hence, no
elementary quanta of the color interaction, in the usual sense,
can exist. This means that no gluon {\it particles} are possible,
and that Eq.(1) does {\it not} hold for color
fields\footnote{Early criticism of this paper focused on that
\textit{physical} particles surely not are as simple as these
``bare" particles, and that by necessity ``dressed" particles must
be used for physical states. This misses the point that the
distinction ``bare" vs. ``dressed" is defined solely within
perturbation theory. Even the ``dressed" particle states rely on
exactly the same kind of approximation methodology as the ``bare"
particles. Also, the classic explanation of the photoelectric
effect uses just these ``naive" quanta, in terms of photons.}. The
{\it fields} are there, but their {\it quanta}, gluons (and
quarks), are relevant only when probed at sufficiently
(infinitely) short distances. Generally, quarks do {\it not}
exchange gluons, but the fermion fields $\psi$ react to the color
fields given by $A_{\mu}$. Fields are primary to particles.

So far we have strictly only banished gluons. To also banish
quarks as physical particles we note that a quark field is the
source of a color field, but this color field is itself a source
of a color field. Hence, a quark field is never removed from other
sources, is always interacting, and can never be considered to be
freely propagating. This results in that the quark fields can
never be represented by harmonic oscillator modes, unless $Q^2
\rightarrow \infty$, whereas \textit{physical} particles,
observable in nature, should exist as $Q^2 \rightarrow 0$. This
means that no quark field quanta (quarks) can ever exist. In QCD,
the particles ``gluons'' and ``quarks'' are merely artifacts of
the approximation method used, {\it i.e.} the perturbative
expansion in the interaction on a ``background'' of {\it assumed}
free gluons and quarks. They are simply absent in the exact
theory.

In QED things are very different. An electric charge gives rise to
an abelian field, which is {\it not} the source of another field.
Hence an electrically charged field {\it can} be removed from
other sources and exist as a physical particle. Thus, the
observability of, {\it e.g.} an electron is ultimately due to the
fact that electromagnetic quanta (photons) can exist as real
particles.

A more mathematical treatment of the physical picture regarding
quarks given above is provided by geometry. A case analogous to
the one we are studying appears in quantum field theory on a
curved spacetime \cite{BirrellDavies}, where it is well known, and
generally accepted, that fields are more fundamental than
particles. Indeed, there it can be shown that the very concept of
a particle is, in general, useless \cite{Davies2}. Actually,
nonabelian gauge fields and quantum field theory on a curved
background have a lot in common. The quark fields can, in an
approximation similar to quantum field theory on a curved
(spacetime) background, be treated as ``living" on the curved
(gauge) space defined by the color fields. The total curvature,
and also the dynamical coupling to ``matter fields'' through the
covariant derivative, is given by one part coming from the
Yang-Mills connection ({\it i.e.} gauge potential) and one part
coming from the Riemannian (Levi-Civita) connection
\cite{NashSen}. Only when {\it both} the gauge field curvature
{\it and} the spacetime curvature \cite{BirrellDavies} are zero,
or at most constant, can a particle be unambiguously defined. The
former is constant for abelian quantum field theory, the latter is
zero on a Minkowski background with inertial observers, and
constant for some special, and static, spacetimes. The curvature
in gauge space is given by the field strength tensor,

\begin{equation}
F_{\mu \nu}^b  =\partial_{\mu} A_{\nu}^b - \partial_{\nu}
A_{\mu}^b + g_s f_{bcd} A_{\mu}^c A_{\nu}^d ,
\end{equation}
this being the analog in gauge space to the Riemann curvature
tensor, $R_{\mu \nu \sigma \rho}$, for spacetime. The properties
of $F_{\mu \nu}$ under a gauge transformation, $U$, is
\begin{equation}
F_{\mu \nu} \rightarrow F_{\mu \nu}' = U F_{\mu \nu} U^{-1},
\end{equation}
{\it i.e.} the gauge curvature generally transforms as a tensor
in gauge space.

However, we see directly that for an abelian gauge theory, like QED,
\begin{equation}
F_{\mu \nu} \rightarrow F_{\mu \nu}' =  F_{\mu \nu},
\end{equation}
as $U$ now commutes with $F_{\mu \nu}$. This means that the
curvature is {\it constant} (invariant) in gauge space for an
abelian field, {\it i.e.} that $F_{\mu \nu}$ transforms as a
scalar in gauge space. It also means that the field is a gauge
singlet, which only reflects that it has no ``charge'' and that
the fields have no self-interactions. Abelian gauge fields $\neq$
sources of fields.

For nonabelian fields, like QCD, the gauge curvature, $F_{\mu
\nu}$ transforms as a tensor, {\it i.e.} is {\it co}variant, not
{\it in}variant, and is thus generally {\it different} at
different points in gauge space. The color-electric and
color-magnetic fields, ${\mathbf{E}}^b$ and ${\mathbf{B}}^b$,
which are the components of $F_{\mu \nu}^b$ defined by $F_{0i} ^ b
= E_i ^b$ and $F_{ij} ^ b = \epsilon_{ijk} B_k ^b$ ($i,j,k \in
1,2,3$), are thus not gauge independent and cannot be observable
physical fields, which is another, complementary and perhaps more
physically direct way of seeing that physical gluons cannot exist,
regardless of coupling strength\footnote{Thus, gluon and quark
``confinement" can be considered as just a special case of the
more general requirement that observables be gauge invariant, {\it
i.e.} independent of the local choice of gauge ``coordinates''.}.
(In contrast to usual electric and magnetic fields which {\it are}
both gauge singlets \textit{and} observed.) The fact that $ F_{\mu
\nu}^b \neq$ color singlet, just implies that color gauge fields
are {\it sources} of color gauge fields.

We conclude that the unbroken nonabelian gauge theories of gravity
and QCD are strictly incompatible with the concept of elementary
quanta in a traditional sense. In practice, however, this only
rules out gluons and quarks as physical particles, as spacetime
curvature (or, equivalently, observer accelerations) is normally
completely negligible in experimental settings in particle
physics. The difference can be traced to the fact that the
dynamical curvature is directly related to the nonlinear coupling
strength, which is enormously much larger for QCD than for
gravity. Leptons can exist as physical particles as QED has
abelian gauge dynamics and weak (nonabelian) SU(2) is broken, {\it
i.e.} absent from the point of view of particle detectors.

It also follows, as a direct corollary to the argument above, that
hadrons must be color singlets, {\it i.e.} color neutral, as they
otherwise could not exist as physical particles.

It would be interesting to continue the analogy with gravity and
speculate that the hadrons are ``grey holes'', as the color stays
inside. The curvature induced by the color fields would then give
the hadron, or confinement, radius.
 This would require nonperturbative solutions to the coupled
$\psi$-$A$ system, with fully dynamical quark fields, which is a
very hard and unsolved problem. Strictly, also gravity should be
included, perhaps in a Kaluza-Klein fashion, the lagrangian then
containing both $F_{\mu \nu} F^{\mu \nu}$, now with covariant
spacetime derivatives, and $R = R^{\mu}_{\mu}$, the Ricci-scalar.
Although this is a nice picture, which may/may not be true, it is
not necessary for the purpose of excluding gluons and quarks as
physical quanta, or particles, for which the argument given in
this article is sufficient.

In conclusion, what we have done is to provide a ``Gordian
knot''-like theoretical explanation of the empirical
``non-appearance" of gluons/quarks in the physical world.
\newline
We assume only that:

1) QCD is the correct theory of quark-field interactions

2) particles (quanta) are represented by $a$ and $a^{\dagger}$
  \newline
which unambiguously leads to the result that QCD can have {\it no}
elementary color quanta (gluons). If a specific fundamental
quantum does not exist within a certain, supposedly correct,
theory it neither can be detected in experiments. As the quark
fields always generate color fields, which in turn act as sources
of other color fields, the quark fields can never be considered to
be noninteracting. Hence an expansion in harmonic oscillator modes
is impossible, which means that no quark field quanta (quarks) can
exist. Only if  I) QCD is wrong, or  II) elementary quanta are
{\it not} necessarily described by harmonic oscillator modes
(which would mean that the fundamental relation $E = \hbar \omega$
does not hold, and that the notion of what a quantum really is
must be broadened from the traditional definition), or both, can
gluons and quarks exist as physical particles. In geometrical
terms the curvature of Yang-Mills (color) space makes quarks, as
particles, impossible.

This proves that the theory of QCD automatically forbids particles
with color charge, hence formally implying gluon/quark
``confinement''. However, in a very real sense, there actually is
nothing to confine in terms of particles. \vspace{0.5cm}
\\{\bf Acknowledgements:} Part of this work was done while
visiting the \textit{Theory Group, Lawrence Berkeley National
Laboratory, Berkeley, CA 94720, USA}. The visit was supported by
the \textit{Foundation BLANCEFLOR Boncompagni-Ludovisi, n\'{e}e
Bildt}.

\end{document}